\documentclass[12pt]{article}

\usepackage{graphicx}

\title{  Chiral physics in the magnetic field with quark confinement contribution }
\author{ M.A.Andreichikov and  Yu.A.Simonov \\
State Research
Center\\Institute of Theoretical and Experimental Physics, \\
Moscow, 117218 Russia}

\newcommand{\beq}{\begin{eqnarray}}
 \newcommand{\eeq}{\end{eqnarray}}
\newcommand{\be}{\begin{equation}}
 \newcommand{\ee}{\end{equation}}

 \def\la{\mathrel{\mathpalette\fun <}}
\def\ga{\mathrel{\mathpalette\fun >}}
\def\fun#1#2{\lower3.6pt\vbox{\baselineskip0pt\lineskip.9pt
\ialign{$\mathsurround=0pt#1\hfil ##\hfil$\crcr#2\crcr\sim\crcr}}}

\newcommand{{\SD}}{\rm SD}

\newcommand{{\Mc}}{\mathcal{M}}

\newcommand{\vex}{\mbox{\boldmath${\rm x}$}}
\newcommand{\vey}{\mbox{\boldmath${\rm y}$}}
\newcommand{\ver}{\mbox{\boldmath${\rm r}$}}
\newcommand{\vesig}{\mbox{\boldmath${\rm \sigma}$}}

\newcommand{\veP}{\mbox{\boldmath${\rm P}$}}
\newcommand{\vep}{\mbox{\boldmath${\rm p}$}}

\newcommand{\vez}{\mbox{\boldmath${\rm z}$}}

\newcommand{\ven}{\mbox{\boldmath${\rm n}$}}

\newcommand{\veB}{\mbox{\boldmath${\rm B}$}}

\newcommand{\lan}{\langle}
\newcommand{\ran}{\rangle}

\begin{document}
\maketitle
\begin{abstract}

The standard chiral perturbation theory is known to  predict much weaker effects in magnetic field, than found in numerical lattice data. To overcome this disagreement we are using 
 the  effective chiral confinement Lagrangian, $L_{ECCL}$, containing both chiral and quark degrees of freedom,  in the  presence of  external magnetic field. Without magnetic  fields $L_{ECCL}$ reduces to the ordinary chiral Lagrangian $L_{EC L}$,  yielding in the lowest order $O(\partial_\mu \varphi)^2$  all known  relations,  and providing  explicit  numerical coefficients  in the higher $O(p^4, p^6)$ orders.
 
 The inclusion of the magnetic field in $L_{ECCL}$  strongly modifies  ECL results for chiral  condensates, coupling  constants $f_\pi, f_K$ and masses of chiral mesons. The  resulting behavior  contains the only parameter -- the string tension $\sigma$, is roughly proportional to  $O\left( \frac{eB}{\sigma}\right)$ and agrees  very well with   lattice data. These results show  that the  magnetic field acts not only on the chiral  degrees of   freedom   $(\varphi_\pi)$, but also on quarks   in the   quark-chiral Lagrangian, which  produce much stronger effect. \end{abstract}

 \section{ Introduction}
 
 The chiral theory \cite{1,2,3} was  developing to a large  extent independently of the theory of quark dynamics and  the chiral mesons  and chiral d.o.f. were treated in a way, which explicitly displayed the flavor symmetry, but relation with quark-antiquark dynamics was obscure. 
 
 In particular,  the leading terms of the effective chiral Lagrangian (ECL) \cite{ 4,5,6} to all orders do not contain any relation to quark dynamics  and quark masses appear actually as correction terms.
 
 Nevertheless  even in this form ECL was very useful to describe low energy reactions with chiral mesons. At higher energies, where internal structure of chiral mesons becomes more important, one needs additional power terms $O(p^4, p^6$) and some modifications of the theory, and the corresponding effective models with adjustable coefficients have been suggested, \cite{7,8,9}, which can describe the  experimental data. 
 
 So far so good, but at some moment an interesting idea has appeared, what could happen, when one  considers the chiral systems in the constant magnetic field (m.f.). As an  immediate reaction it was suggested, that one can use as an effective Lagrangian in m.f. the well known ECL with the replacement of the derivatives $(\partial_\mu U)^n$ in its terms by the full  derivative $D_\mu U = \partial_\mu U + ie A_\mu   U $ which takes into account the action of m.f. on the chiral d.o.f. in $U=\exp\left( \frac{i \hat \varphi_a \lambda_a}{f_\pi}\right)$.

This idea was considered well substantiated and it was used in many papers, both in the ECL \cite{10,11,12,13,14,15,16} and in the NJL model \cite{18,19,20}.

In particular, for the quark condensate in \cite{10} with the definition  $\lan \bar q q\ran \equiv \sum = | \lan \bar u u \ran | = |\lan \bar d d \ran |$  it was found, using chiral perturbation theory (ChPT)
\be \sum(B) = \sum (0) \left[ 1+\frac{eB ln 2}{16 \pi^2 F^2_\pi} + O\left( \frac{(eB)^2}{F^4_\pi}\right) \right]. \label{1}\ee

Note the integer value of charge $e$ in (\ref{1}), and the same linear behavior for $\lan \bar u u\ran$ and $\lan \bar d d \ran$ condensates, which is the consequence of purely chiral degrees of freedom, having only integer charges, but looks unrealistic for the quark observable with  charge $e_q\neq e$.
In \cite{10} also the relation for $F_{\pi^0}
$ and $M_{\pi_0}$ were found, supported later with  corrections due to $m^2_\pi\neq 0$  in \cite{11,12,13,14,15,16,17}
\be \frac{F^2_{\pi^0}(B)}{F^2_{\pi^0}( 0)} = 1 + \frac{eB ln 2}{ 8\pi^2 F^2_\pi}+... \label{2}\ee

 \be  {M^2_{\pi^0}(B)}= M^2_{\pi^0} (0)\left[ 1-\frac{eB ln 2}{16 \pi^2 F^2_\pi} + ...\right]  \label{3}\ee
 The appearance  of the factor $1/F^2_\pi$ in (\ref{1}), (\ref{2}), (\ref{3}) is not  surprising, since this is the basic  dimensional factor in the ECL.
 \be L= \frac{F^2_\pi}{4} tr (\partial_\mu U \partial_\mu U^+) + O(\hat m_q U)+...\label{4}\ee

 The idea of  a  specific chiral-magnetic physics, mostly independent of  separate quark degrees of freedom, was and still is attractive and produced several directions.  
 One of this is the theory suggested in \cite{21}.
 
 One assumes in this type of approach that quark d.o.f. are inessential for low values of $eB\ll (4\pi F_\pi)^2$, where the phenomena  of the chiral magnetic system can be observed, while  for higher m.f. as was argued in \cite{10} the $q\bar q$ d.o.f. can be important, but due to asymptotic freedom quarks can be considered as free of strong interaction and only subject to m.f. Thus one can neglect strong interactions, and first of all the confinement, considering chiral systems, made of quarks, in weak and strong m.f. However, the accurate lattice calculations of the quark  condensate $|\lan \bar q q\ran|$ in \cite{22,23,24,25} have provided much stronger linear growth of  $|\lan \bar q q\ran|$, than in the ChPT \cite{10}. In general, the numerous lattice data in \cite{22,23,24,25,26,27,28,29,30,31, 31*} demonstrate much stronger influence of m.f. on chiral observables, than is predicted by ChPT.
 
 It is a purpose of this paper to present a theory and explicit calculations which can explain this disagreement. We take into account quarks and antiquarks with confinement  in the framework of the Effective Chiral Confinement Lagrangian (ECCL) \cite{32,33,34,35,36,37,38,39,40} in m.f. and  demonstrate, that confinement yields much larger effects due to m.f., than in the purely chiral theory shown  in (\ref{1}),(\ref{2}),(\ref{3}).
 
 We also show that our results are in good agreement with available lattice data \cite{22,23,24,25,26,27,28,29,30,31, 31*} for the same physical quantities. 
 
 To understand the importance of confinement in the magnetic effects of chiral observables one can compare the pure chiral results (\ref{1}),(\ref{2}),(\ref{3}), which can be written as $X_{\Sigma, F, M} = 1 + a_{\Sigma, F, M}^{(ChPT)} \frac{eB}{(4\pi F_\pi)^2} $  $+...$, where $ a_{\Sigma, F, M} =O(1)$,  which means that chiral observables are essentially  controlled by the parameter  $(4\pi F_\pi)^2 \sim O(1$ GeV$^2)$.
 
  This is to  be compared with the effects of m.f. with the  confinement interaction between $q$ and $\bar q$, which has the order of magnitude $O\left( \frac{eB}{\sigma} \right)$,  with the standard string tension  $\sigma =0.18$ GeV$^2$.  In  particular, for neutral pion decay constant $F_{\pi^0}$ one obtains,   $\left( \frac{F_{\pi^0} (eB)}{F_{\pi^0} (0)}\right)=  \sqrt{1+\left(\frac{e_qB}{\sigma}\right)^2}$.
  
  Moreover,  the confinement is the basic interaction which allows to calculate all coefficients in the ECL including the orders $O(p^4)$ and $O(p^6)$  as it is shown in \cite{36}.
  
  Therefore one expects that the theory of chiral observables based on both quark and chiral d.o.f.  should provide a more  appropriate set up for the  calculation of all quantities $\Sigma, F, M$ for  both neutral and charged pions in m.f. are in good agreement with lattice data \cite{22,23,24,25,26,27,28,29,30,31,31*}. As will be seen, our  formalism provides numerical results  for all observables $\Sigma, F, M$ without fitting parameters in the few percent  agreement with all lattice data.

  The paper is organised as follows. In the next section we start from the ECCL Lagrangian and  derive all three chiral observables $\Sigma, F, M$ without  m.f. taking into account both chiral and quark d.o.f.   We pay below a special attention to the separation of purely $q\bar q$ and chiral d.o.f. and discuss also higher order corrections.  In section 3 we discuss the  effects of m.f. on chiral and  confinement dynamics  and obtain  $\Sigma, F, M$ in the presence of m.f. In section 4 we shortly demonstrate the results of ECCL calculations for $\Sigma , F, M$  in m.f.and compare those with lattice data. In section 5  a short conclusion and discussion is  given.

  \section{The Effective Chiral Confinement Lagrangian}

  The main idea of the quark-chiral or Effective Chiral Confinement Lagrangian  (ECCL), approach \cite{32,33,34,35,36} is to take into account  simultaneously pure chiral d.o.f. and quark d.o.f. connected first of all with confinement. The corresponding Lagrangian, called in \cite{36} the Effective Chiral Confinement Lagrangian was derived in  \cite{32,33,34,35,36} in  the following form
  \be L_{eff} (M_s, \hat \phi) = - N_c tr {\rm log} [ i \hat \partial + \hat m + M \hat U].\label{5}\ee where $\hat U= \exp (\hat \phi \gamma_5)$, and $M(x)$ is a confining kernel, which represents confinement of the quark with the mass $m_i$ (here $\hat m= {\rm diag} (m_i))$, in the common Wilson loop of the $q\bar q$ Green's function, as shown in Fig.1. One can take $M(x) = \sigma |\vex|$ as shown in Fig.1, where it is convenient to take $\vex =0$ at the midpoint of quark and antiquark. In what follows one encounters $M(x)$ at the vertex of the $q\bar q$ Green's function, which  we associate with the minimal distance $x_{\min} = \lambda \cong0.1$ fm, and we keep $M(0) = M(\lambda) = \sigma \lambda= 0.15$ GeV as the only parameter of  our Lagrangian, besides $\sigma$ and   current quark masses $m_i$. It is understandable that with magnetic field (m.f.) one defines  $\hat \partial \to \hat \partial - ie \hat A^{(e)}$.


\begin{figure}
\includegraphics[width= 13cm,height=14cm,keepaspectratio=true]{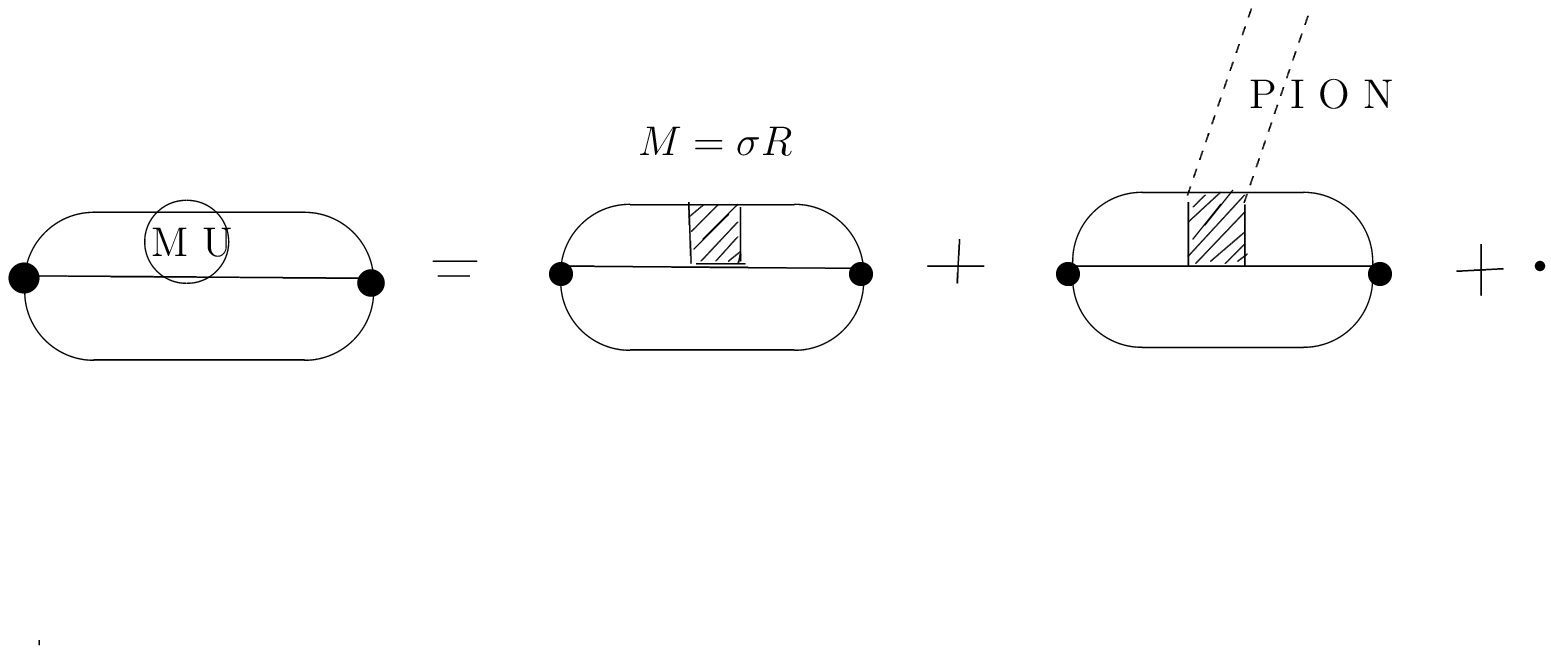}

\caption{ The effective chiral confining kernel $M_s  \hat U$ comprising
confinement in $M_s$ and chiral mesons in $ \hat U$. } \vspace{1cm}

\end{figure}

  
  To proceed and to find the connection with the standard ECL \cite{1,2,3,4,5}, one defines the quark Green's function \be S_0 = \frac{i}{\hat \partial + \hat m + M} \equiv i \Lambda\label{6}\ee
  and 
  \be \eta = \hat U^+ \Lambda (\hat \partial + \hat m ) ( \hat U - 1) = \eta_\varphi + \eta_m\label{7}\ee
  with 
  \be \eta_\varphi = \hat U^+ \Lambda\hat \partial \hat U, ~~ \eta_m = \hat U^+ \Lambda \hat m (\hat U-1),\label{8}\ee
  so that $L_{eff}$ acquires the form 
  \be L_{eff} =- N_c tr {\rm log} (1-\eta).\label{9}
  \ee
  
 As a result the  quadratic in $\hat \partial \hat U$ part of Lagrangian is

\be L_{eff}^{(2)}  =\frac12  N_c tr ( \hat U^+ \Lambda\hat \partial \hat U, \hat U^+ \Lambda\hat \partial \hat U) \approx  \frac{N_c}{2} tr  ( \Lambda\hat \partial \hat\varphi \bar \Lambda \hat \partial \hat\varphi),\label{10}
  \ee with $\bar \Lambda = \hat m +M -\hat \partial, ~ \hat \varphi = \frac{\varphi_i \lambda_i}{F_i},~ F_i = F_\pi, F_K$
   are numbers.
   
   Here the sign $tr$ implies summation over flavor, Dirac and space-time coordinates.
   
   To understand connection of $L_{eff}^{2)}$ in (\ref{10})  with the standard  ECL Lagrangian (\ref{4}), one can express  $\Lambda$ via the quadratic quark Green's function $G$ as 
   \be \Lambda=(m+M-\hat \partial) G.\label{11}\ee
   
   As a result one obtains 
   \be L_{eff}^{(2)} = \left\{ \frac{\hat f^2_\pi}{4} (\partial_\mu U \partial_\mu U^+)\right\}, \label{12}\ee where the operator  $ \hat f^2_\pi$ can be written  in terms of Hamiltonian eigenvalues.

For the latter one can exploit the path integral form, suggested in \cite{41,42,43} and repeatedly used for all mesons and baryons. In terms of the n-th wave function $\varphi_n (r)$ of the quark $a$ and antiquark $b$ and eigenvalue $M_n$, as well as $\omega^{(n)}_{a,b} = \lan \sqrt{\vep^2 + m^2_a}\ran_n$, the chiral decay constant is \cite{36,33,40,41, 42}
\be \left( f^{(n)}_{ab}\right)^2 = \frac{N_c (m_a+ M(0))(m_b+ M(0))}{2\omega_a^{(n)} \omega_b^{(n)} M_n \xi_n}\varphi_n^2(0),\label{13}\ee
where $\xi_n$ for light quarks is $\xi_n = 1/2.34$ \cite{42,43}. 

The good accuracy of (\ref{13}), supported by  comparison of calculations in \cite{40, 41,43} with experiment  \cite{44},  allows to use the chiral quark theory of ECCL in the case of external m.f.

   Note the difference between numbers $F_i, F_k$ and operators $\hat f^2_{ik}$, which contain derivatives $\hat\partial$ also in $G_i, G_k$ and hence can depend on  external magnetic field. In absence of m.f. the expression (\ref{13}) was used  to calculate the decay contains of $F_\pi, F_K, F_D$   in the framework of the path integral Hamiltonian \cite{45,46,47}. As a result  one obtains the physical eigenvalues of $\hat f^2_{ik}$, which agree well with  experimental values as shown in \cite{41,42,43}. Finally, choosing these values as parameters $F_i$ in $\hat \varphi$, one accomplishes the standard ECL Lagrangian $L_{ECL}^{(2)}$,where $F_i$ serve as basic  dimensional parameters of the theory. 
   
   In our  case in  the ECCL Lagrangian these parameters can be calculated in terms of basic QCD parameters: string tension $\sigma, \alpha_s$ and current quark masses $m_i$. 
   
 Using $M(0)=0.15$ GeV and $\sigma = 0.18$ GeV$^2$ one obtains in \cite{43} for $\bar f_\pi = \sqrt{2} f_\pi, ~\bar f_k =\sqrt{2} f_k$  the following values 
 \be\bar f_\pi = 0.133~{\rm GeV},~~  \bar f_k =  0.165~{\rm GeV},\label{14}\ee
 which should be compared with experimental values \cite{44}
 \be\bar f_{\pi^+}^{(exp)}  = (130.7 \pm 0.1 \pm 0.36)~{\rm MeV},\label{15}\ee
 
\be\bar f_{k^+}^{(exp)}  = (159.8 \pm 1.4 \pm 0.44)~{\rm MeV}.\label{16}\ee

In a  similar way one computes $f_\pi,f_K$ for higher radial excitations, as it is shown in \cite{43}.

First of all one can check the GMOR relations \cite{2} which follow from the ECCL in the second order $O(\eta^2)$.
\be L_{ECCL}^{(2)} = N_c tr \left(\eta + \frac{\eta^2}{2}\right) = N_c tr \left\{ \Lambda m \frac{\hat \varphi^2}{2} + \frac12 \Lambda \hat \partial \hat\varphi \bar \Lambda \hat \partial \hat \varphi - \frac12 \Lambda \hat m \hat\varphi \bar \Lambda \hat m \hat \varphi\right\}.\label{19}\ee

Taking into account that 
\be N_c tr \Lambda \hat m \frac{\hat \varphi^2}{2} = \Delta_a m_a \varphi_{ab} \varphi_{ba}  \label{20}\ee 
one obtains, neglecting $O(m^2)$ terms from the last term in (\ref{19}),
\be f^2_{\pi^0} M^2_{\pi^0} = \Delta_{1}m_1  + \Delta_{2}m_2 \label{21}\ee
\be f^2_{\pi^+} M^2_{\pi^+} =f^2_{\pi^-} M^2_{\pi^-} =\Delta_{1}m_1  + \Delta_2m_2 \label{22}\ee
\be f^2_{K^+} M^2_{K^+} =f^2_{K^-} M^2_{K^-} =\Delta_{1}m_1  + \Delta_3m_3 \label{23}\ee
\be f^2_{K^0} M^2_{K^0} =\Delta_{2}m_2  + \Delta_3m_3. \label{24}\ee Here $\Delta_a$ is the quark condensate, $\Delta_a = N_c tr \Lambda_a$, 
where 1,2,3 refer to $u,d,s, $ and $m_i$ are pole quark masses, which are connected to the current quark masses in $\overline{MS}$ scheme (see \cite{44} and \cite{47}). One can see that (\ref{21})--(\ref{24}) coincide with the standard GMOR relations \cite{2}. 

Considering now the quark condensate $\Delta_a$ one has $$ \Delta_a = N_c \lan tr \Lambda_a\ran = N_c \left\lan Tr \left( \frac{1}{\hat \partial + m_a+M} ( M+ m_a- \hat \partial) \frac{1}{M+m_a - \hat \partial}\right) \right\ran = $$
\be = N_c (m_a + M(0)) Tr (\Lambda_a \bar \Lambda_a) = N_c (m_a + M(0)) Tr (\Lambda_a \gamma_5 \Lambda_a\gamma_5).\label{25}\ee

To proceed one can express $\Delta_a$ via the Green's function $G_{aa}(0)$,   in \cite{33,34}, which was exploited in \cite{37} in the case of nonzero m.f., so that $\Delta_a=N_c (m_a +M(0)) G_{aa}(0)$  
\be G_{ab } (k) = \int d^4 (x-y) e^{ik (x-y)} \lan Tr     \Lambda_a  (x,y) \gamma_5 \Lambda_a(y,x)\gamma_5)\ran .\label{26a}\ee

Using the spectral  decomposition of $G_{ab}(k)$,

\be G_{ab}(k)= \sum^\infty_{n=0}   \frac{c_n}{k^2+ \bar M^2_n},\label{27}\ee
one finds 
\be \Delta_a = N_c (M(0) + m_a) \sum^\infty_{n=0} \frac{\varphi^2_n (0)}{M_n }.\label{26}\ee

Here $\varphi_n(r), M_n$ are eigenfunction and eigenvalue of the Green's function $G_{ab}$ and the corresponding Hamiltonian,  found in \cite{46,47,48}, which correspond to $J^{PC} = 0^{-+}$ and do not contain chiral d.o.f.  In a similar way from $G_{ab} (k) = G_{ab} (0) + \frac{k^2f^2_\pi}{N_c} +...$ one obtains as  in \cite{33,34} for $f_{ab}$. 
\be f_{ab}^2 =N_c (M(0) + m_a) (M(0) + m_b) \sum^\infty_{n=0} \frac{|\varphi_n (0)|^2}{M^3_n}\label{27a}\ee
 Note, that  forms (\ref{13}) and  (\ref{27a}) are equivalent, since for the PS states $\varphi_n$ $2 \omega^2_a \xi_n \approx M^2_n , ~~ M_n \approx 2 \omega_a (n)$.

In what follows we shall mostly use the forms (\ref{26}), (\ref{27a}). To proceed one must detalize   the Hamiltonian technic, which produces $M_n,  $   $\varphi_n(0)$   to prepare for the inclusion of m.f. in this Hamiltonian.

Note, that  confinement is separated in $\Lambda_a = \frac{1}{\hat \partial + m+ M(x)}$ as the interaction term $M(x) = \sigma |\vex |$, which implies the use of the instantaneous interaction $V_{q\bar q} (\vex - \vey ) = M(\vex) + M_{\bar q} (\vex)$,  between $q$ and $\bar q$.  This form was useful above to introduce together with confinement the chiral d.o.f. as $M(\vex) \to M(\vex) U(x)$. 

To calculate effects of confinement  and all spin corrections in our case of $
G_{ab} (x,y)$, where chiral d.o.f. do not participate, it is more  convenient to go back to origial QCD Green's function with confinement, as it is done in \cite{42, 43}. 

Now using the path integral representation in the Euclidean space-time with the proper time $s_i = \frac{T_4}{2\omega_i}, ~ T_4 = x_4-y_4$, one has as in \cite{42,43}
\be I_{ab} \equiv\left( \frac{1}{(m^2_a- \hat D^2_a) (m^2_b- \hat D^2_b)}\right)_{xy} = \frac{T_4}{8\pi} \int^\infty_0 \frac{d\omega_a}{\omega_a^{3/2}}  
 \int^\infty_0 \frac{d\omega_b}{\omega_b^{3/2}}(D^3z_a)_{\vex\vey}(D^3z_b)_{\vex\vey} e^{-A}\label{28}\ee
 where \be A=K_a (\omega_a) + K_b(\omega_b)+ \int^{T_4}_0 dt_E V_0 (r(t_E))\label{29}\ee
 and $V_0(r)$ is the result of the instantaneous interaction from the Wilson loop
 \be V_0(r) = V_{\rm conf} (r) + V_{OGE} (r) +\Delta V, \label{30}\ee
 where $\Delta V$ includes spin-depend  part, and 
 \be K_i = \frac{m_i^2 + \omega^2_i}{2\omega_i} T_4 + \int^4_0 dt_E \frac{\omega_i}{2} \left( \frac{d\vez^{(i)}}{dt_E} \right)^2, ~~ i=a,b.\label{31}\ee
 
 From (\ref{28})--(\ref{31}) one obtains in a  standard way the Hamiltonian
 
\be I_{ab}= \frac{T_4}{8\pi} \int^\infty_0 \frac{d\omega_a}{\omega_a^{3/2}}  
 \int^\infty_0 \frac{d\omega_b}{\omega_b^{3/2}}\left\lan \vex, \vex \left| e^{-H(\omega_a, \omega_b, \vep_a, \vep_b) T_4} \right| \vey, \vey\right\ran,\label{32}\ee
 and 
 \be H= \sum_{i=a,b} \frac{(\vep^{(i)})^2 + m^2_i + \omega^2_i}{2\omega_i} + V_0 (r) = \frac{\veP^2}{2(\omega_1+ \omega_2)} + \frac{\vep^2}{2\tilde\omega^2} + V_0(r) \equiv \frac{\veP^2}{(2 (\omega_1+\omega_2)}+h.\label{33}\ee

 As a result the matrix element $\lan | e^{-H T_4}|\ran$ in (\ref{32}) can be written as
 \be \int d^3 (\vex-\vey) \left\lan \vex, \vex \left| e^{-H T_4} \right| \vey, \vey\right\ran = \sum_n \varphi^2_n (0) e^{-M_n (\omega_a, \omega_b) T_4},\label{34}\ee 
 where $\varphi_n, M_n$ are eigenvalues of $h\equiv \frac{\vep^2}{2\tilde\omega} + V_0(r), $ $\tilde \omega= \frac{\omega_a\omega_b}{\omega_a+ \omega_b}$, $h\varphi_n = M_n \varphi_n $. 
 
 In the limit $T_4 \to \infty$ one can use the stationary point analysis of the integral $\int I_{ab} d^3 (\vex-\vey)= \int^\infty_0 \frac{d\omega_a}{\omega_a^{3/2}} \int^\infty_0 \frac{d\omega_b}{\omega_b^{3/2}} \varphi^2_n (0) e^{-M_n(\omega_a, \omega_b) T_4}$, which finally yields the stationary values $\omega_a^{(0)}, \omega_b^{(0)}$ from the  condition $\frac{\partial M_n}{\partial\omega_i} |_{\omega_i = \omega_0^{(0)}} = 0, i=a,b$, and the final physical eigenvalues $M_n^{(0)} = M_n (\omega_i^{(0)})$ and eigenfunctions $\varphi^{(0)}_n = \varphi_n (\omega_i^{(0)}, \ver)$.

 As it was shown in \cite{33,34} the resulting eigenvalues of the (nonchiral) PS states with confinement, color Coulomb and  spin-spin interactions taken  into account, are $M_0 =0.4$ GeV, $M_1=1.35$ GeV and $M_2 =1.85$, while $|\varphi_n (0)|^2 = \frac{\omega_n^{(0)}}{4\pi} \left( \sigma+ \frac43 \alpha_s \lan \frac{1}{r^2}\ran \right)$.
 
 Finally, using (\ref{27a}) one  obtains the value of $f^2_{\pi^0}\cong f^2_{u\bar u} \cong f^2_{d\bar d} = 94$ MeV, where the first 3 states of the fast  converging series in (\ref{27a}) are taken into account. This value agrees well with (\ref{14}), obtained in \cite{43} in a different way. 
 
 For $\Delta_a$ the same values of $M_n, |\varphi_n (0)|^2$ can be taken but the series is formally diverging and must be  renormalized. We shall not touch this point below,  since we shall need the difference $\frac{\Delta_a (e_aB) - \Delta_a (0)}{\Delta_a (0)}, $ where only few first terms contribute.  
 
  In the next section we generalize this derivation  imposing m.f. on our $q\bar q$ system. 
 
 \section{The $q\bar q$  system in magnetic field}
 
 In this case the Hamiltonian can be  written  as \cite{49,50}
 \be H= \sum_{i=a,b} \left( \frac{(\vep_i- e_q \bar A^{(e)})^2}{2\omega_i} + \frac{m^2_i+\omega^2_i}{2\omega_i} - \frac{e_i \vesig_i \veB}{2\omega_i} \right) + V_0 (R) \label{35}\ee
 and for the neutral systems the relative  motion  Hamiltonian is 
 \be h=\frac{1}{2\tilde \omega} \left( - \frac{d^2}{d\ver^2} + \left( \frac{e_q (\ver \times \veB)}{2}\right)^2 \right)+ V_0 (r)+\sum_{i=a,b}\frac{e_i \vesig_i \veB}{2\omega_i} \label{36}\ee 
  
 An analytic answer for energy eigenvalue  and eigenfunctions can be obtained with $O(5\%)$ accuracy replacing the linear  confinement by  the  quadratic form with the  subsequent  stationary point  analysis of coefficients 
 \be V_{\rm conf}^{(\rm lin)} = \left.  \sigma r \to \frac{\sigma}{2} \left( r^2 \gamma + \frac{1}{\gamma}\right), ~~ \frac{\partial M}{\partial \gamma}\right|_{\gamma=\gamma_0} =0, \label{37}\ee
 which yields the final result
 \be M_{\ven} = \varepsilon_{\ven} + \frac{m^2_a+\omega_a^2 - e_a\veB \vesig_a}{2\omega_a} + \frac{m^2_b+\omega_b^2 - e_b \veB \vesig_b}{2\omega_b}.\label{38}\ee
 
 Here $\varepsilon_{\ven}$ is 
  \be \varepsilon_{\ven}=\frac{1}{2\tilde \omega} \left( \sqrt{e^2_q B^2+ \sigma^2 c} (2n_\bot +1) + \sqrt{\sigma^2 c} \left(n_3 +\frac12\right) \right) + \frac{\gamma \sigma}{2} \label{39}\ee
  and $c=\frac{4\tilde \omega}{\gamma\sigma}$, while $\gamma\to \gamma_0$ is defined from the minimum of $M_{\ven} (\gamma)$. 
  
  The index $\ven$ here denotes $n_\bot , n_3$ and two possible relative orientation of $\vesig_a, \vesig_b$ with respect to $\veB$, $(+-)$ and $(-+)$.
  
  The most important role in what follows is played by the factor  $|\varphi_n (0)|^2$ in (\ref{26}), (\ref{27a}), which is easily calculated in the oscillator potential to be 
  \be |\varphi_n (0) |^2 = \frac{1}{\pi^{3  /2} r^2_\bot r_3}, ~~ r^2_\bot = 2 ((e_q B)^2 + \sigma^2c)^{-1/2}, ~~r_3 =\left(\frac{\sigma^2c}{4} \right)^{-1/4}.\label{40}\ee
  
  Defining $c=c_{+-}, c_{-+}$ by stationary point analysis one find that 
  \be c_{+-} (B) \approx 1, ~~ c_{-+} (B) \cong \left(1+ \frac{8e_q B}{\sigma} \right)^{2/3}\label{41}\ee
  and as a result one has \cite{37}
  \be |\psi^{(+-)}_{n_\bot =0,n_3} (0) |^2 \cong \frac{\sqrt{\sigma} \sqrt{ e^2_q B^2 + \sigma^2}}{(2\pi)^{3/2}}\label{42}\ee
 
  \be |\psi^{(-+)}_{n_\bot =0,n_3} (0) |^2 =(\sigma^2 c_{-+})^{3/4} \sqrt{1+ \left( \frac{e_q B}{\sigma} \right)^2 \frac{1}{c_{-+}}}.\label{43}\ee

It is interesting, that the energy $M_{\ven}$ of  the lowest level with $n_\bot = n_3 =0$ is very  different in the $(+-)$ and $(-+)$ cases. Indeed $M^{(+-)}_0$ is slowly  changing and tends to the constant limit, while $M^{(-+)}_0$, where the  cancellation of the terms $e_i \vesig_i \veB$ does not take place, is growing with $eB$:
\be  M^{(-+)}_0 (a,a)\approx 2 \sqrt{2|e_aB|}, ~~ M_0^{(+-)} \cong  const. \label{44}\ee

We now can write expressions for $(\Sigma, F, M)$ in m.f. in a general form, generalizing Eqs. (\ref{26}), (\ref{27a}) since now the sum over $n$ is going over $n_\bot, n_3$ and $(+-), (-+)$, so that one has \cite{38, 39}\footnote{Note an  erroneous  dependence of $M^{(+-)}(eB)$ on m.f.   in \cite{39} and  as a result  a  fast growth of $f_{\pi^0} (eB)$. In what  follows we shall not use these results.} 
\be f^2_{\pi^0} (e_q B) = N_c ( M(0) + m_i )^2 \sum_{n_\bot, n_3} \left( \frac{ \frac12 |\psi_{ni}^{(+-)} (0)|^2}{(M_{ni}^{(+-)})^3}+ \frac{ \frac12 |\psi_{ni}^{(-+)} (0)|^2}{(M_{ni}^{(-+)})^3}\right),\label{45}\ee
where $i=u,d$. In the same way $\Delta_a$ acquires the form \cite{37,38}
  \be \Delta_i = N_c |\lan \bar q_iq_i\ran | = N_c ( M(0) + m_i )  \sum_{n_\bot, n_3} \left( \frac{ \frac12 |\psi_{ni}^{(+-)} (0)|^2}{ M_{ni}^{(+-)}  }+ \frac{ \frac12 |\psi_{ni}^{(-+)} (0)|^2}{ M_{ni}^{(-+)}  }\right).\label{46}\ee
  
  Finally, it was found in \cite{38}, that GMOR relations \cite{2} for neutral mesons are  conserved in m.f., so that we can write
  \be m^2_{\pi^0} f^2_{\pi^0} = \frac{\bar m M(0)}{M(0) +\bar m} |\lan \bar u u\ran + \lan \bar d d \ran |, ~~\bar m = \frac{m_u + m_d}{2}\label{47}\ee
  and one can use GMOR relations  separately for $u$ and $d$ flavors. In the next section we shall find analytically and numerically the behavior of $\Delta_i$, $f_{\pi^0}$ and $m_{\pi^0}$ in m.f.

  \section{Behavior of chiral observables $(\Sigma, F, M)$ in m.f}

 One can define explicitly the behavior of  
  $ \Sigma, F, M $ in m.f. introducing the relative growth coefficients
  \be \frac{\Delta_a (e_aB) - \Delta_a (0)}{\Delta  (0)} \equiv  \Delta \Sigma_a , ~~a = u,d.\label{48}\ee
  \be \frac{ f^2_{\pi^0} (e_a B)}{f^2_{\pi^0}(0)} = K^{(a)}_f (B)\label{49}\ee
  \be \frac{ m^2_{\pi^0} (  B)}{m^2_{\pi^0}(0)} = K_{\pi^0 } (B).\label{50}\ee
  
  One can notice in (\ref{45}), (\ref{46}), that masses $M_{ni}$ in the  denominator strongly increase, when one is going from $n_\bot, n_3 =0,0$ to higher values and there occurs a significant  compensation of higher $n_\bot, n_3$ terms in the difference $\Delta_a (B) - \Delta_a (0)$ etc. Therefore the main  contribution to the coefficients $ \Delta\Sigma_a, K_f, K_\pi$ is given by the  lowest term with $n_\bot, n_3=0,0$, which we retain and compare to the lattice data. 
  
  For $\Delta\Sigma_a(B)$ one obtains from (\ref{46}), (\ref{42}), (\ref{43})\be \Delta \Sigma_a = \frac12 \left\{ \sqrt{1+\left( \frac{e_a B }{\sigma}\right)^2}+\sqrt{1+\left( \frac{e_a B }{\sigma}\right)^2 \frac{1}{c^{(a)}_{-+}}} - 2 \right\}, ~~ a=u,d.\label{51}\ee 
  
  For $K_f^{(a)} (e_aB)$ one can use (\ref{44}), (\ref{45}) and neglect   the terms $(-+)$ in (\ref{45}), since $M^{(-+)}_n$ is fast growing at large $eB$. As a result one obtains a simple estimate
  \be 
  K_f^{(a)} (e_aB) \cong \sqrt{1+\left( \frac{e_a B }{\sigma}\right)^2}
  , ~~ e_a B \ga\sigma \label{52}\ee
  
  Finally for $m^2_{\pi^0}$ one has form (\ref{47})

  \be \frac{ m^2_{\pi^0} (e_a  B)}{m^2_{\pi^0}(0)} = \frac{\Delta_u (e_uB)} 
   { f^2_{\pi^0} (e_u B)}\frac{f^2_{\pi^0}(0)}{\Delta_u(0)}= \frac{1+ \Delta\Sigma_a}{ K^{(a)}_f (e_aB)}.\label{53}\ee
   
   We start with the quark condensate, Eq. (\ref{51}) and define two functions in analogy with \cite{25}:
   \be \frac12 (\Delta \Sigma_u+ \Delta \Sigma_d)\equiv K_+(eB)\label{54}\ee
   \be  \Delta \Sigma_u- \Delta \Sigma_d \equiv K_-(eB)\label{55}\ee
   
   In Fig.2,3 we show both functions (\ref{54}), (\ref{55})  computed with the help of (\ref{51}) in comparison with  the chiral perturbation theory result (\ref{1}), and  with the lattice calculated ratios in  \cite{25}. One can see a very good agreement of our result with lattice data for all measured $eB \leq 1.1$ GeV$^2$.  One notice two important distinctions of our and lattice  results with the ECL results
\begin{figure}
\includegraphics[width= 13cm,height=14cm,keepaspectratio=true]{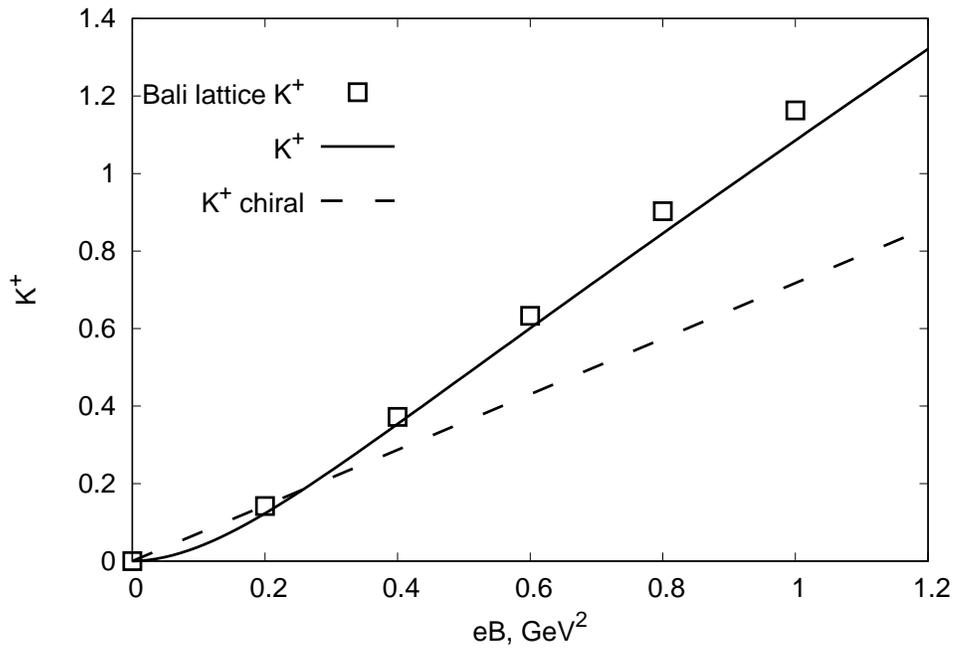}

\caption{ The $K_+(eB$ calculated with ECL (solid line) in comparison with the lattice data from  \cite{25} and chrial perturbation theory (dashed line).} \vspace{1cm}

\end{figure}
\begin{figure}
\includegraphics[width= 13cm,height=14cm,keepaspectratio=true]{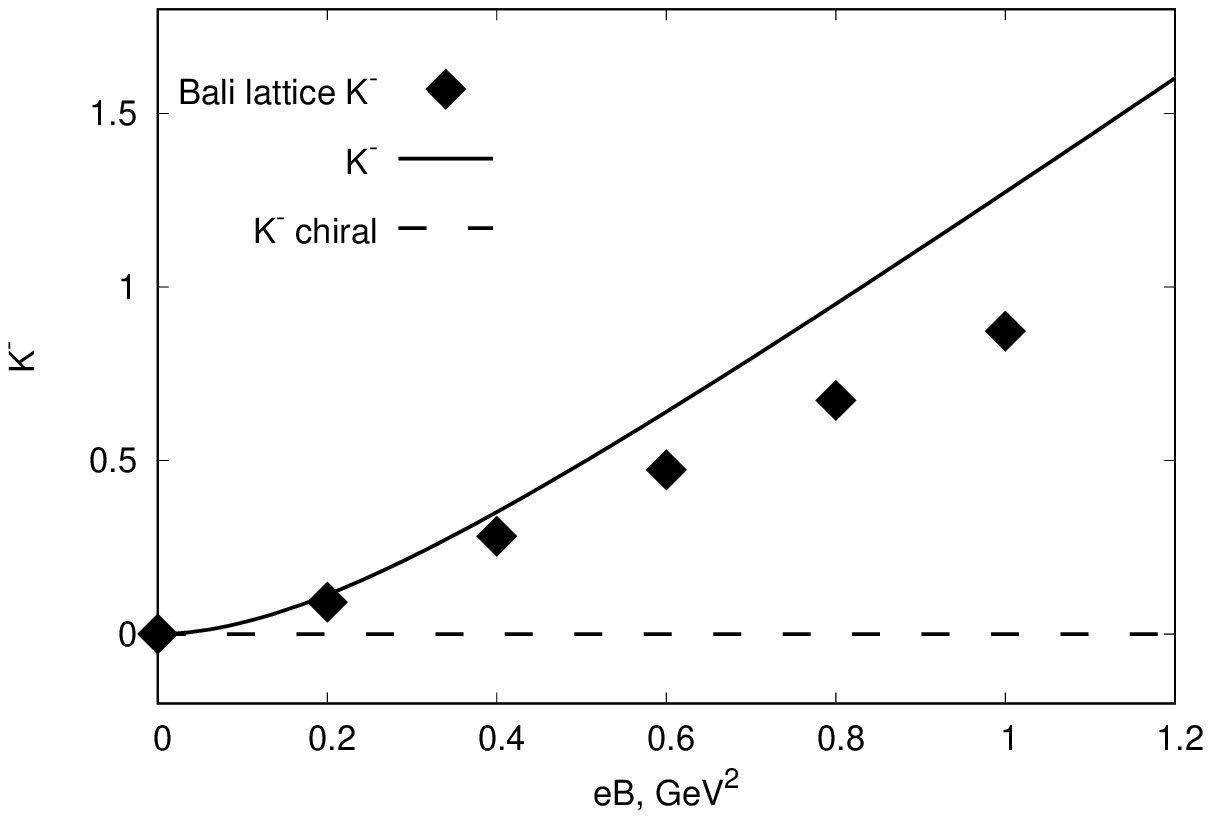}

\caption{ The $K_{-}(eB$ calculated with ECL (solid line) in comparison with the lattice data from  \cite{25} and chrial perturbation theory (dashed line). } \vspace{1cm}

\end{figure}

   \begin{enumerate}
   \item $\Delta \Sigma_q$ is roughly proportional to $|e_qB|$ at large $e_q B\ga \sigma$, and hence $\Delta \Sigma_u \approx 2 \Delta \Sigma_d,$ while in (\ref{1}) $ \Delta \Sigma_u = \Delta \Sigma_d$, so that in CHPT $K_-(eB) \equiv 0$.  
   \item The asymptotic linear behavior of  $\frac12 (\Delta \Sigma_u+ \Delta \Sigma_d)$ from (\ref{51}), (\ref{54}) is $\left( \frac{eB}{4\sigma}\right)=\frac{1.4 eB}{1{\rm GeV}^2},$ and is given by the $(+-)$ states,  which agrees very well numerically with lattice data \cite{25}, while in ChPT  $K_+ \cong \frac{0.5 eB}{1{\rm GeV}^2}$.
   \end{enumerate}
   
   We now turn to the case of $f^2_{\pi^0}$ and display in Fig. 4 our result (\ref{53}) in comparison with the ChPT result, Eq. (\ref{2}). One can see different behavior of these two results. In our case the asymptotics is given in (\ref{52}), \\ $ K_f^{(a)} \cong \frac{e_aB}{\sigma} \approx  \left(  3.7, 1.85 \right) \frac{eB}{1{\rm GeV}^2}$, for $(u,d)$  while in ChPT result (\ref{2}), this ratio is $\frac{ln 2 eB}{8\pi^2 F^2_\pi} \cong  0.25 \frac{eB}{1{\rm Gev}^2}$.
   \begin{figure}
\includegraphics[width= 13cm,height=14cm,keepaspectratio=true]{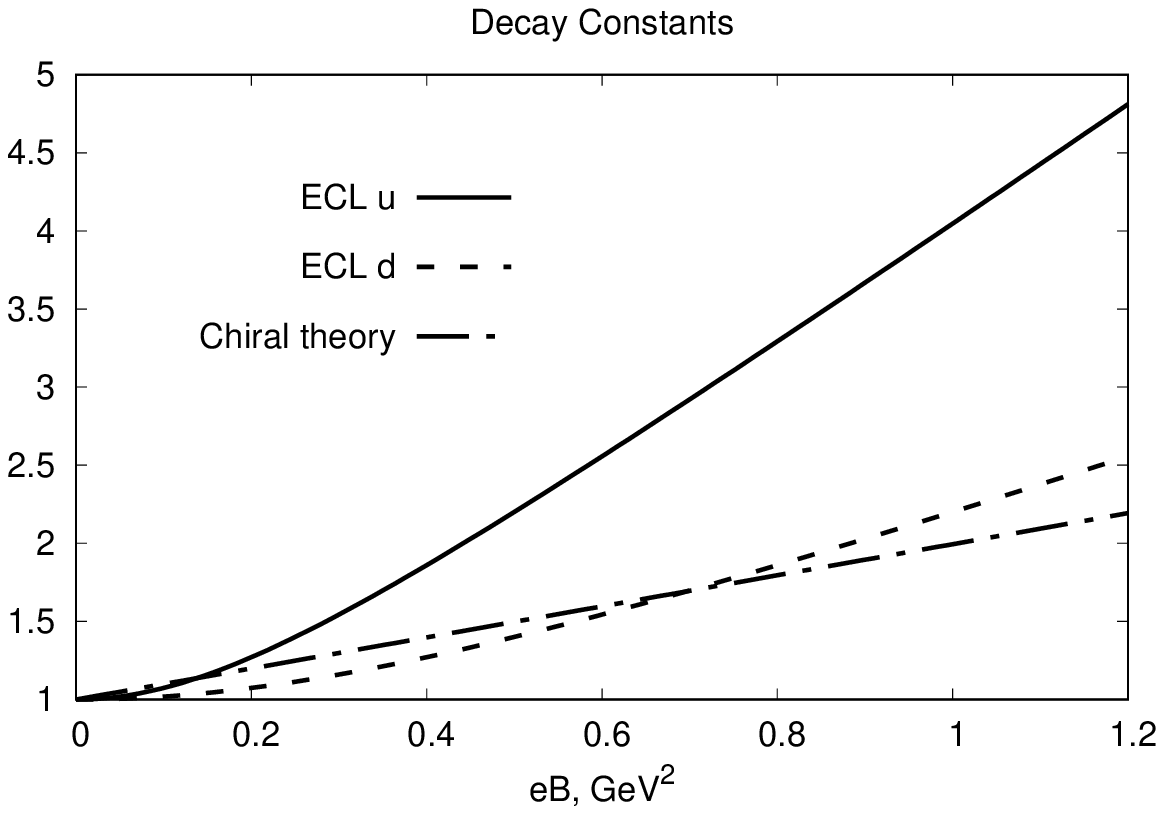}

\caption{ The decay constants $\frac{f_{\pi^0}^2(eB)}{f_{\pi^0}^2(0)}$ for $u \bar{u}$ (solid) and $d \bar{d}$ (dashed) quark constiuents calculated with ECL in comparison with the standard chiral perturbation theory. } \vspace{1cm}

\end{figure}

   We finally come to the $\pi^0$ mass problem, which according to (\ref{53}) can be written as 
   
   \be \frac{ m^2_{\pi^0} (e_q  B)}{m^2_{\pi^0}(0)} = \frac{1+A}{2}\label{56}\ee
   with \be A= \left[ \frac{1+\left( \frac{e_q B }{\sigma}\right)^2 \frac{1}{c_{-+}}}  {1+\left( \frac{e_a B }{\sigma}\right)^2  }  \right]^{1/2} .\label{57}\ee 
   
   The resulting curves for $q=u,d$ a shown in Fig.5,6 together with the lattice data from \cite{31*,31} and the ChPT result Eq. (\ref{3}) from \cite{10}. One can see a reasonable  agreement of our result with lattice data \cite{31,31*} and again a strong diasagreement with Eq. (\ref{3} ).
   \begin{figure}
\includegraphics[width= 13cm,height=14cm,keepaspectratio=true]{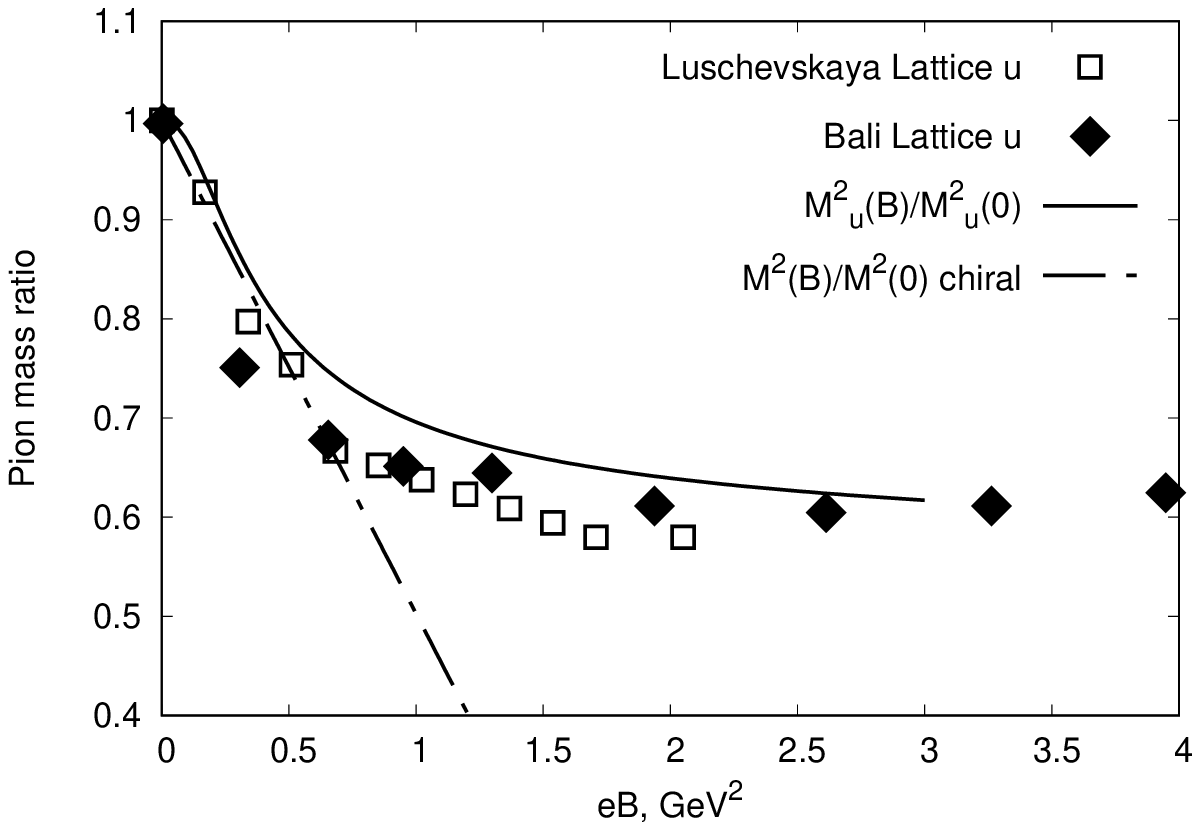}

\caption{ The pion mass ratio $\frac{m_{\pi^0}^2(eB)}{e_{\pi^0}^2(0)}$ with $u \bar{u}$ quark contituents calculated with ECL (solid line) in comparison with the lattice data from \cite{31*,31} (dots) and with the standard chiral perturbation theory prediction from \cite{10}. } \vspace{1cm}

\end{figure}
\begin{figure}
\includegraphics[width= 13cm,height=14cm,keepaspectratio=true]{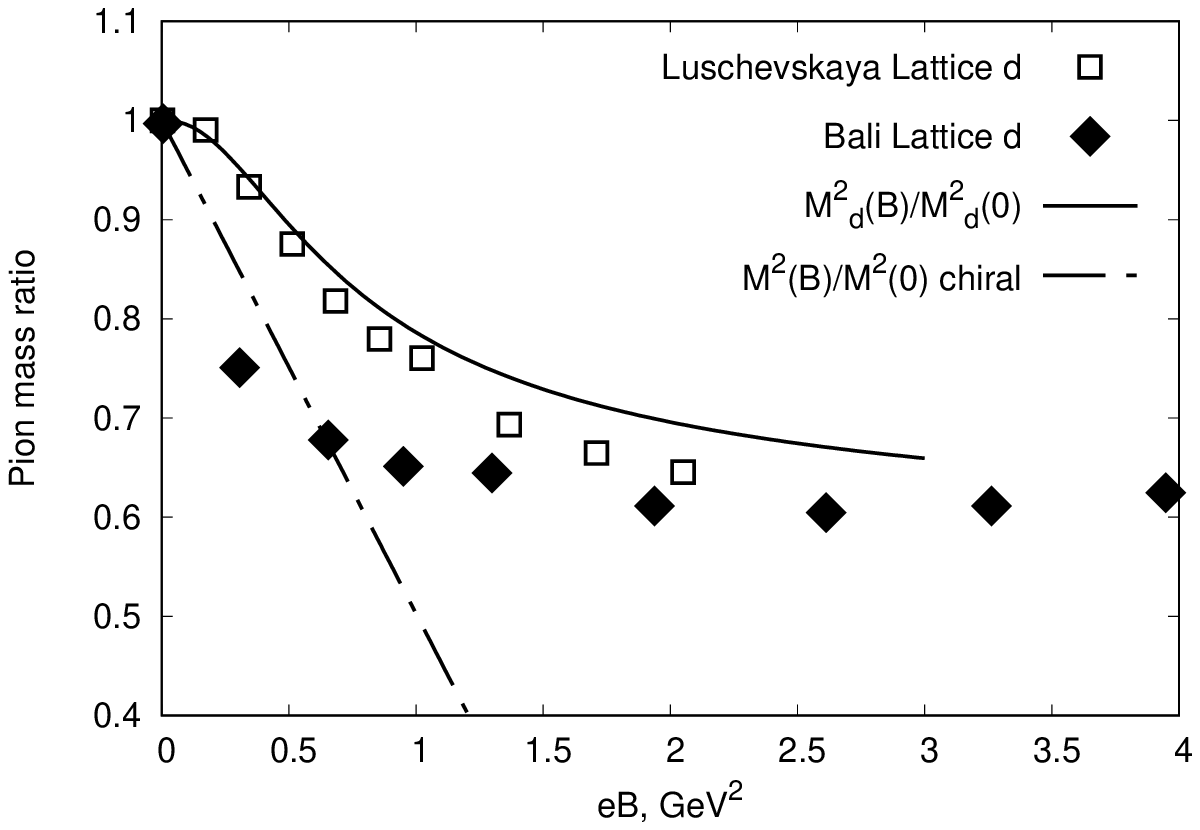}

\caption{ The pion mass ratio $\frac{m_{\pi^0}^2(eB)}{e_{\pi^0}^2(0)}$ with $d \bar{d}$ quark contituents calculated with ECL (solid line) in comparison with the lattice data from \cite{31*,31} (dots) and with the standard chirla perturbation theory prediction from \cite{10}. } \vspace{1cm}

\end{figure}
  
\section{Results and discussion}

 Our results for $\Sigma,F, M$ are shown in Figs. 2-6  in comparison with lattice data and results of CHPT, Eq. (\ref{1}), (\ref{2}), (\ref{3}). One can see  a good agreement of our results in Fig.2,3  and Fig.5,6  with  lattice data from \cite{25} and \cite{31,32} respectively, while in  all Figs. 2-6 an apparent  disagreement with the ChPT, Eqs. (\ref{1}), (\ref{2}), (\ref{3}). One should stress that our eqs. for  ($\Sigma,F, M$) do not contain any  fitting parameters and depend only on $\frac{e_q B}{\sigma}$, where $\sigma=0.16$ GeV$^2$ is the standard QCD string tension. 
 
 It is understandable, that in our approach the basic effect is from the quark d.o.f (as  it seen in the  coefficients $\frac{eB}{\sigma}$), and  the agreement with independent lattice data shows that this effect is properly taken into account. One of the points is then: where is the contribution from purely chiral d.o.f.? This point is especially  aggravated, when one compares the coefficient $K_-(eB)$, (\ref{55}) in Fig.3, which according to ChPT should be identically zero, while in our and lattice approaches it is essentially nonzero and is of the same order, as the total ChPT correction to $\Delta (eB)$.
 
  In our derivation of $\Delta_a$ and $\Delta \Sigma_a$, Eqs. (\ref{26}), (\ref{48}) the  chiral corrections are absent and they appear in ECCL only in higher orders, in the  terms $O(\Lambda(\hat\partial\varphi)^n), n> 2$. The same  happens in the case of $f_{\pi^0}, M_{\pi^0}$, where to the lowest order the terms $(\partial_\mu \pi^\pm )^2$ are absent.

  As a general feature, one should stress, that the ECCL is nonlocal in chiral variables and e.g. the pion propagator is replaced by the $q
  \bar q$ Green's function with pion quantum numbers, therefore the purely local chiral magnetic effects can be expected only for very small momenta and m.f.

  We have not considered above the charged mesons $\pi^\pm, K^\pm$, where GMOR relations are violated in the lowest order,as it was found in \cite{38}. The corresponding mass evolution for $M_{\pi^\pm}$ is given in \cite{40}.
  
  One can compare our and lattice results with other approaches beyond ChPT. In \cite{25} the lattice results for $K_+$ have been compared to the PNJL model of \cite{52}, showing a reasonable agreement for $eB\la 0.3$ GeV, and   deviation from lattice data for larger $eB$. A better agreement of $K_+$ with lattice data  in the whole  interval $eB\leq 1$ GeV$^2$ was found in \cite{53}, in the framework of the NJL model with a Gaussian formfactor. The $\pi^0$ mass in the NJL model, $M_{\pi^0} (eB)$, was found numerically in \cite{54,55,56}. One can see in Fig. 2 of \cite{56} almost the same slope, as in our Fig.5,6, slightly different (within 15\%) from the lattice data of \cite{32} and within errors with data of \cite{31}. A good agreement of \cite{53} Fig.4, with our data can be found for $f^2_{\pi^0}(eB)$ in Fig.4. These  coincidences support the main outcome of our paper, that the quark d.o.f., taken into account also within the NJL model, play the most important role in the  impact of m.f.
  
  It is interesting to identify the explicit mechanism, which provides the linear growth of $\Sigma_a (eB)$, and $f_{\pi^0} (eB)$ with increasing $eB$. Indeed, looking at Eqs. (\ref{42}), (\ref{43}) and (\ref{45}), (\ref{46}) one can notice, that the main effect of increase comes from the factor $|\varphi(0)|^2 \sim \sqrt{1+ \left( \frac{e_qB}{\sigma}\right)^2}$, which is  a familiar effect of the $q \bar q$ attraction at small distances in m.f., called in \cite{57} the ``magnetic focusing  effect''. This  effect is present both in relativistic and nonrelativistic systems. One can notice that it is  specially important in the spin-spin interactions, providing collapse  of the $q\bar q$ system in the lowest local approximation of $hf$ interaction in m.f. This point was treated in \cite{50,58}, where it was shown that an effective smearing is necessary for spin-spin forces in m.f., which prevents collapse and satisfies the positivity conditions for eigenvalues. This kind of  treatment is also assumed in our case.
  
  It is possible, that the  agreement of our results   with the corresponding data from \cite{53, 56} is due to the same simple magnetic focusing effect, discussed above. 
  
  Having  found that purely chiral ECL Lagrangian, not containing quark d.o.f., does not ensure the correct behavior of the physical system under the influence of m.f., one may ask, what  happens to the so-called chiral magnetic  effects in similar purely chiral Lagrangians?  This question requires a detailed analysis and a possible extension of these purely chiral Lagrangians to the quark-chiral form, as it is done in the extension of $L_{ECL}$ to $L_{ECCL}$ in \cite{36}.
  
  This work was supported by the Russian Science Foundation grant 16-12-10414.

 \end{document}